\documentclass[a4paper,12pt]{article}
\usepackage[pctex32]{graphics}
\usepackage{amssymb,amsmath}
\usepackage{amsmath,amssymb}
\usepackage{latexsym}
\usepackage{epsfig}
\usepackage[english]{babel}

\newcommand{\be}{\begin{equation}}
\newcommand{\ee}{\end{equation}}
\newcommand{\ba}{\begin{eqnarray}}
\newcommand{\ea}{\end{eqnarray}}

\begin{document}

\begin{titlepage}

\vspace{5mm}
\begin{center}
{\Large \bf Quasinormal modes and hidden conformal symmetry in the
Reissner-Nordstr\"om black hole}

\vskip .6cm

\centerline{\large
 Yong-Wan Kim $^{1,a}$,  Yun Soo Myung$^{2,b}$,
and Young-Jai Park$^{1,3,c}$}

\vskip .6cm

{$^{1}$ Center for Quantum Spacetime,
Sogang University, Seoul 121-742, Korea}\\

{$^{2}$Institute of Basic Science and School of Computer Aided
Science, \\Inje University, Gimhae 621-749, Korea \\}

{$^{3}$Department of Physics and Department of Global Service Management,\\
Sogang University, Seoul 121-742, Korea}

\end{center}

\begin{center}

\underline{Abstract}
\end{center}
It is shown that the scalar wave equation in the near-horizon limit
respects a hidden SL(2,R) invariance in the Reissner-Nordstr\"om
(RN) black hole spacetimes.  We use the SL(2,R) symmetry to
determine algebraically the  purely imaginary quasinormal
frequencies of the RN black hole. We confirm that these are exactly
quasinormal modes of scalar perturbation  around the near-horizon
region  of a near-extremal black hole.

\vskip .6cm

\vskip 0.8cm

\vspace{15pt} \baselineskip=18pt

\noindent $^a$ywkim65@gmail.com\\
\noindent $^b$ysmyung@inje.ac.kr \\
\noindent $^c$yjpark@sogang.ac.kr

\thispagestyle{empty}
\end{titlepage}

\newpage
\section{Introduction}

It was known that  the scalar wave equation in the near-region and
low-energy limits enjoys a hidden conformal symmetry  in the
non-extremal Kerr black hole which is not  an underlying symmetry of
the spacetime itself~\cite{Castro:2010fd,Krishnan:2010pv}.  The
existence of such a hidden symmetry originates from the observation
that scattering amplitudes of scalar off a black hole are given in
terms of hypergeometric functions \cite{Cvetic:1997xv,Cvetic:2011hp}
which form representations of the conformal group SL(2,R).
Importantly, this led to the conjecture that the non-extremal Kerr
black hole with angular momentum $J$ is dual to a CFT$_2$ with the
central charges $c_L=c_R=12J$~\cite{Castro:2010fd}, which provides
exactly the Bekenstein-Hawking entropy of the Kerr black hole.  It
is also found that  the low energy scalar-Kerr scattering amplitudes
coincide with thermal correlators of a CFT$_2$.

Chen and Long~\cite{Chen:2010ik} have shown that one can use the
hidden conformal symmetry to algebraically determine quasinormal
mode spectrums as descendants of a highest weight state in black
hole spacetimes. On the other hand,  the spin-2 and spin-3
quasinormal modes and frequencies around the BTZ black hole were
constructed by the purely operator approach without any
approximation~\cite{SS,MKMP,Myung:2012sh}.

Recently, the authors~\cite{Bertini:2011ga} have shown that the
scalar wave equation in the near-region and  low-energy limits
enjoys a hidden SL(2,R) invariance in the Schwarzschild geometry.
They have used the SL(2,R) symmetry to determine algebraically the
quasinormal frequencies (QNFs) of the Schwarzschild black hole, and
also shown that this yields the purely imaginary QNFs describing
large damping. Explicitly, starting with the highest weight state
$\Phi^{(0)}$ with $L_0\Phi^{(0)}=h\Phi^{(0)}$ and $L_1\Phi^{(0)}=0$,
all quasinormal modes could be constructed as descendants of
$\Phi^{(n)}=(L_{-1})^n \Phi^{(0)}$ obtained  by acting with $L_{-1}$
on the highest weight state $\Phi^{(0)}$. Then, one can read off the
QNFs from the descendants.

We would like to mention that the method developed for the Kerr/CFT
correspondence could not be directly applied to the Schwarzschild
and RN black holes because there is no apparent AdS$_2$ structure in
the near-horizon geometry of the non-extremal Schwarzschild and RN
black holes. In this direction, a hidden conformal symmetry could be
extracted by making a five-dimensional  uplifted RN black
hole~\cite{Chen:2010as}. However, this is not a genuine conformal
symmetry developing in the four-dimensional RN black hole. Very
recently, the authors~\cite{Ortin:2012mt} have found a hidden
SL(2,R) symmetry in the near-region and low-energy limits of the RN
black hole.

In this work, we will use the SL(2,R) symmetry to determine QNFs of
the RN black hole algebraically. Employing  the operator method, we
derive quasinormal modes (\ref{qnms}) and  purely imaginary QNFs
(\ref{impq}). A key point in deriving the quasinormal modes is to
set the proper boundary conditions for the wave equation.  As is
well known, quasinormal modes are  determined  by solving a scalar
wave equation around the RN black hole as well as imposing the
boundary conditions: ingoing waves at the horizon and outgoing waves
at infinity of asymptotically flat spacetime.   It seems difficult
to derive QNFs of a scalar propagating on the RN  black hole by
using a hidden conformal symmetry solely. The reason is that
quasinormal modes do not satisfy the outgoing wave-boundary
condition  because these modes were developed using the near-region
and low-energy limits~\cite{Bertini:2011ga} where the frequency
$\omega$ should satisfy inequalities of $\omega \ll 1/r$ and $\omega
\ll 1/r_+$.  Importantly, the geometry modified in this way (the
subtracted geometry) shows that it has the same near-horizon
properties as the original RN black hole, but different asymptotes
~\cite{Cvetic:2011dn}.  The latter is given by the asymptotically
anti-de Sitter (AdS) spacetime.  That is, developing the hidden
conformal symmetry in the near-horizon is necessary to make change
of  the boundary condition at infinity  from asymptotically flat to
asymptotically AdS spacetime.   It is suggested that quasinormal
modes (\ref{qnms}) satisfy the ingoing-boundary condition at the
horizon and Dirichlet boundary condition at infinity.

However, it requires  an appropriate picture described by the
solution to a scalar wave equation around a specified RN black hole
to derive QNFs.  We should  find  the specified RN black hole which
captures quasinormal modes (\ref{qnms}) and QNFs (\ref{impq}). It is
found that {\it the specified RN black hole is given by  the
near-horizon region of a near-extremal RN black hole}.  The wave
equation (\ref{sch3}) corresponds to (\ref{sch4}) around the AdS
segment of AdS$_2\times S^2$, which is just the geometry of the
near-extremal RN  black hole~\cite{Li:2011tga}.  We show in Appendix
that the purely imaginary QNFs (\ref{impq}) could be found exactly
by solving the scalar wave equation  around the near-horizon region
of near-extremal RN black hole.  Also, quasinormal modes
(\ref{qnmhg}) as the solution to the scalar wave equation around the
near-horizon region of a near-extremal RN black hole satisfy the
ingoing-boundary condition at the horizon and Dirichlet boundary
condition at infinity. Even the authors in~\cite{Chen:2012zn} have
used the different boundary condition, we obtain the same QNFs
(\ref{neqnfs}) when taking  the massless and neutral limits.
Finally, we will also discuss QNFs by adapting potential pictures.

\section{Hidden conformal symmetry}

First, let us introduce the RN black hole whose metric is given by
 \be\label{MF}
 ds^2_{\rm RN}=
  -f(r)dt^2+f^{-1}(r)dr^2+r^2d\theta^2+r^2\sin^2\theta d\phi^2
 \ee
with the metric function
 \be
 f(r)=1-\frac{2M}{r}+\frac{Q^2}{r^2}.
 \ee
Here, $M$ and $Q$ are the  ADM mass and the electric charge of the
RN black hole, respectively.  Then, the inner ($r_-$) and the outer
($r_+$) horizons are obtained as
 \be
 r_\pm=M\pm\sqrt{M^2-Q^2}\equiv M\pm r_0,
 \ee
which satisfy $f(r_\pm)=0$. We note that $r_0$ is a non-extremal
parameter, but a very small $r_0 \ll M(\sim Q)$ corresponds to the
near-extremal RN black hole.  Also,  we have an extremal RN black
hole for $r_0=0$.

For the RN black hole, the relevant thermodynamic quantities are the
Bekenstein-Hawking entropy and Hawking temperature
 \ba\label{entropy}
 S_{BH}&=& \pi r_+^2,\\
 \label{temp}
 T_{H}&=&\frac{r_+ - r_-}{4\pi r_+^2}=\frac{r_0}{2\pi r_+^2},
 \ea
respectively. Note that the surface gravity is defined as
 \be\label{surg}
 \kappa=\frac{r_0}{r_+^2}=2\pi T_H.
 \ee

Now, let us consider a minimally coupled massless scalar propagating in
the spacetimes (\ref{MF}), which satisfies the Klein-Gordon equation
 \be \label{KG}
 \bar{\square}_{\rm RN}\Phi = 0.
 \ee
Using the ansatz
 \be
 \Phi(t,r,\theta,\phi) = e^{-i\omega t}\frac{R(r)}{r}Y^l_m(\theta,\phi)\,
 \ee
together with the eigenvalue equation on $S^2$
 \ba
 \Delta_{S^2}Y^l_m(\theta,\phi) &=&
   \frac1{\sin\theta}\partial_{\theta}(\sin\theta\partial_{\theta}Y^l_m(\theta,\phi))
 + \frac1{\sin^2\theta}\partial^2_{\phi}Y^l_m(\theta,\phi)\nonumber\\
 &=& -l(l+1)Y^l_m(\theta,\phi),
 \ea
the Klein-Gordon equation (\ref{KG}) transforms into the
Schr\"odinger equation
 \ba \label{sch1}
 \frac{d^2}{dr_*^2} R(r) +\Big[\omega^2-V_{\rm RN}(r)\Big]R(r)=0.
 \ea
Here the tortoise coordinate is defined by $dr_*=dr/f(r)$ and the
potential is given by
 \be  \label{pot1} V_{\rm RN}(r)=f(r)\Big[\frac{l(l+1)}{r^2}+\frac{2M}{r^3}-\frac{2Q^2}{r^4}\Big].
 \ee

Next, we  consider a coordinate transformation of
 \be\label{rho}
 \rho \equiv -\frac{1}{2\kappa}\ln\Big[1-\frac{2r_0}{r-r_-}\Big].
 \ee
In terms of $\rho$, the event horizon $r=r_+$ is mapped into
$\rho\rightarrow\infty$, while the spatial infinity
$r\rightarrow\infty$ into $\rho\rightarrow 0$: $r\in[r_+,\infty]$ is
mapped to $\rho\in [\infty,0]$. In this work, we will only consider
outside the event horizon because we wish to compute quasinormal
modes. For the interior of the Cauchy horizon, see
Ref.~\cite{Ortin:2012mt}.  Using the new coordinate (\ref{rho}), the
RN metric~\cite{Ortin:2012mt,Ferrara:1997tw} becomes
 \be\label{metric}
 ds^2_{\rm \rho} =
  -\tilde{f}(\rho)dt^2+\tilde{f}^{-1}(\rho)\left(\frac{r_0}{\sinh(\kappa \rho)}\right)^2
  \left[\left(\frac{\kappa}{\sinh(\kappa \rho)}\right)^2  d\rho^2
  +d\theta^2+\sin^2\theta d\phi^2\right]
 \ee
with
 \be
 \tilde{f}(\rho)
 =\frac{1}{\left(e^{\kappa\rho}+\frac{r_-}{r_0}\sinh(\kappa\rho)\right)^{2}}.
 \ee
Here  we note a useful relation between $r$ and $\rho$
 \be\label{req}
 r^2=\tilde{f}^{-1}(\rho)\left(\frac{r_0}{\sinh(\kappa\rho)}\right)^2.
 \ee

Then, let us consider a minimally coupled massless scalar
propagating in the spacetimes (\ref{metric}), which satisfies the
Klein-Gordon equation
 \be \label{rhoKG}
 \bar{\square}_{\rm \rho}\Phi = 0.
 \ee
Using the ansatz
 \be\label{ansatz}
 \Phi(t,\rho,\theta,\phi) = e^{-i\omega t}R(\rho)Y^l_m(\theta,\phi)\ ,
 \ee
the Klein-Gordon equation (\ref{rhoKG}) transforms into the
second-order differential equation expressed in terms of  $\rho$
 \be\label{radial}
 \Big(\frac{\sinh(\kappa\rho)}{\kappa}\Big)^2\frac{d^2}{d\rho^2}R(\rho)
 +\left[\frac{\omega^2}{\tilde{f}^2(\rho)}\Big(\frac{r_0}{\sinh(\kappa\rho)}\Big)^2
 - l(l+1)\right]R(\rho)=0,
 \ee
which is again transformed into the Schr\"odinger-type
equation~\cite{Daghigh:2006hu}
 \ba
 \label{sch2}
 \frac{d^2}{d\rho^2} R(\rho) +\Big[\omega^2-V_{\rm \omega}(\rho)\Big]R(\rho)=0.
 \ea
Here the $\omega$-dependent  potential is given by
 \be \label{pot2}
 V_{\rm \omega}(\rho)=
 \omega^2\Big[1-\frac{(\kappa r_0)^2}{\tilde{f}^{2}(\rho)\sinh^4(\kappa\rho)} \Big]
 +\frac{l(l+1)\kappa^2}{\sinh^2(\kappa\rho)}.
 \ee

In order to develop a hidden conformal symmetry,  we  introduce
three vector fields
 \ba\label{vecops}
 L_1 &=& \frac{1}{\kappa} e^{\kappa t}
       \Big[\cosh\left(\kappa\rho\right)\partial_t+\sinh\left(\kappa\rho\right) \partial_\rho\Big]\ , \nonumber \\
 L_0 &=& -\frac{1}{\kappa}\partial_t \ , \\
 L_{-1} &=& \frac{1}{\kappa}e^{-\kappa t}
       \Big[\cosh\left(\kappa\rho\right)\partial_t-\sinh\left(\kappa\rho\right) \partial_\rho\Big]\  , \nonumber
 \ea
which  are slightly different from the previous
construction~\cite{Ortin:2012mt}.
 These satisfy  the SL(2,R) commutation relations
 \be
  \left[L_0, L_{\pm 1}\right]=\mp L_{\pm 1},
  ~~~\left[L_1, L_{-1}\right]=2L_0.
 \ee
Then, the SL(2,R) Casimir operator is constructed  by
 \ba\label{H2}
 {\cal H}^2&=&L^2_0-\frac{1}{2}(L_1L_{-1}+L_{-1}L_1)\nonumber\\
           &=&-\Big(\frac{\sinh(\kappa\rho)}{\kappa}\Big)^2\partial^2_t
              +\Big(\frac{\sinh(\kappa\rho)}{\kappa}\Big)^2\partial^2_\rho   \ .
 \ea
We approximate the second term in Eq. (\ref{radial})  as
 \ba \label{app1}
 && \frac{\omega^2}{\tilde{f}^2(\rho)}\left(\frac{r_0}{\sinh(\kappa\rho)}\right)^2
 =\omega^2 r^4 \left(\frac{\sinh(\kappa\rho)}{r_0}\right)^2
 \\
 \label{app2} && \approx \omega^2 (r_++2r_0 e^{-2\kappa\rho})^4
 \left(\frac{\sinh(\kappa\rho)}{r_0}\right)^2 \\
 \label{app3}
 &&\approx \omega^2 r_+^4 \left(\frac{\sinh(\kappa\rho)}{r_0}\right)^2=\omega^2\left(\frac{\sinh(\kappa\rho)}{\kappa}\right)^2,
 \ea
where we use (\ref{req}) to obtain  (\ref{app1}), and  use the
near-horizon approximation of $r\approx r_++2r_0e^{-2\kappa \rho}$
  to derive (\ref{app2}). Finally, we obtain
(\ref{app3}) by employing the low-energy
approximation~\cite{Bertini:2011ga}. In the near-region and
low-energy approximations which are necessary to develop the hidden
conformal symmetry, the first two terms of the $\omega$-dependent
potential (\ref{pot2}) disappear, leading to the last term only.

As a result, comparing  (\ref{radial}) with (\ref{H2}), the
Klein-Gordon equation in the near-region and low-energy
approximations can be rewritten in terms of the SL(2,R) Casimir
operator ${\cal H}^2$ as
 \be\label{casi0}
 \bar{\square}_{\rm \rho}\Phi = 0 \to
  {\cal H}^2\Phi=l(l+1)\Phi. \ee
The latter can be rewritten as the Schr\"odinger equation
 \be \label{sch3}
 \frac{d^2}{d\rho^2} R(\rho) +\Big[E-V_{\rm HCS}(\rho)\Big]R(\rho)=0,
 \ee
where the energy $E$ is
 \be
 E=\omega^2,
 \ee
and  the HCS-potential takes the form
 \be\label{pot3}
 V_{\rm HCS}(\rho)=\frac{l(l+1)\kappa^2}{\sinh^2(\kappa\rho)}.
 \ee
Therefore, the massless scalar wave equation carries the hidden
conformal symmetry which is not a spacetime symmetry. We note  that
the hidden conformal symmetry is realized  only when approximating
the $\omega$-dependent potential $V_{\rm \omega}(\rho)$ (\ref{pot2})
by the HCS-potential $ V_{\rm HCS}(\rho)$ (\ref{pot3}) in the
Schr\"odinger equation. See Fig. 1-(b) for their difference.

At this stage we would like to emphasize that in the near-horizon
limit $(\rho\to \infty)$ the HCS-potential takes a form of $V_{\rm
HCS}\sim e^{-2\kappa \rho}$, while at infinity $(\rho\to 0)$ it goes
to a form of $V_{\rm HCS}\sim \frac{1}{\rho^2}$. It shows that
$V_{\rm HCS}(\rho)$ is similar to the potential of a scalar field
around the AdS-black hole.  This may imply  that  its asymptote is
changed from a flat spacetime implied by the RN black hole to an AdS
spacetime.

\section{Quasinormal modes constructed by operator method}

We are now in a position to  use the hidden conformal symmetry to
derive QNFs of the RN black hole.  First, we define the primary
state by $\Phi^{(0)}$ which satisfies
 \be
 L_0\Phi^{(0)}=h\Phi^{(0)},
 \ee
and the highest weight condition
 \be\label{hwc}
 L_1\Phi^{(0)}=0.
 \ee
Since $\Phi^{(0)}$ takes the form
 \be
 \Phi^{(0)}=e^{-i\omega_0 t}R^{(0)}(\rho)Y^l_m(\theta,\phi),
 \ee
one has a conformal weight
 \be\label{h1}
 h=i\frac{\omega_0 }{\kappa}=i \frac{\omega_0}{2\pi T_H}.
 \ee
On the other hand, for  $\Phi^{(0)}$, the SL(2,R) Casimir
operator satisfies
 \be \label{caop}
 {\cal H}^2\Phi^{(0)}=h(h+1)\Phi^{(0)}.
 \ee
Comparing Eq. (\ref{caop}) with Eq. (\ref{casi0}), one has
 \be
 h=\frac{1}{2}[1\pm(2l+1)].
 \ee
Together with Eq.~(\ref{h1}), one can find
 \be
 \omega_0=-i\frac{\kappa}{2}[1\pm(2l+1)].
 \ee
Proposing that  the QNFs are purely imaginary
$\omega_I<0~(\omega=\omega_R+i\omega_I)$ with $\omega_R=0$, we
choose the upper sign as
 \be
 \omega_0=-i\kappa(l+1).
 \ee
Then,  all the descendants are constructed  by
 \be
 \Phi^{(n)}=(L_{-1})^n\Phi^{(0)}
 \ee
so that we have
 \be
 \Phi^{(n)}=e^{-i\omega_n t}R^{(n)}(\rho)Y^l_m(\theta,\phi),
 \ee
where the QNFs are read off as
 \be \label{impq}
 \omega_n=\omega_0-i\kappa n=-i\kappa\Big[n+l+1\Big],
 \ee
which is our main result.

Moreover, the $n$-th radial eigenfunction $ R^{(n)}(\rho)$ takes the form
 \ba
 R^{(n)}(\rho)&=&\left(\kappa\right)^{-n}
    \left(-i\omega_{n-1}\cosh\left(\kappa\rho\right)-\sinh\left(\kappa\rho\right)\frac{d}{d\rho}\right)\nonumber\\
    &&~\times\left(-i\omega_{n-2}\cosh\left(\kappa\rho\right)-\sinh\left(\kappa\rho\right)\frac{d}{d\rho}\right)\nonumber\\
    &&~\cdot\cdot\cdot \times \left(-i\omega_0\cosh\left(\kappa\rho\right)-\sinh\left(\kappa
    \rho\right)\frac{d}{d\rho}\right)R^{(0)}(\rho). \label{qnms}
 \ea
We also have
 \be
 L_0\Phi^{(n)}=(h+n)\Phi^{(n)},
 \ee
which implies that $\Phi^{(n)}$  forms a principal discrete highest
weight representation of the SL(2,R). Now  we wish to  solve the
highest weight condition (\ref{hwc}) to determine the highest weight
state $R^{(0)}(\rho)$
 \be
 \Big[-i\omega_0\cosh\left(\kappa\rho\right)+\sinh\left(\kappa\rho\right)\frac{d}{d\rho}\Big]R^{(0)}(\rho)=0.
 \ee
The solution is given by
 \be\label{R0sol}
 R^{(0)}(\rho)=C\Big[\sinh\left(\kappa\rho\right)\Big]^{i\frac{\omega_0}{\kappa}}.
 \ee
Here we note that the tortoise coordinate $r_*$ given by
 \be
 r_*=r-\frac{r^2_-}{2r_0}\ln(r-r_-)+\frac{1}{2\kappa}\ln(r-r_+)
 \ee
approaches
 \be
 r-r_+ \sim e^{2\kappa r_*},~~{\rm as}~~ r\rightarrow r_+.
 \ee
On the other hand,  $\rho$-coordinate (\ref{rho}) goes to
 \be
 r-r_+ \sim e^{-2\kappa\rho}
 \ee
in the near-horizon region so that  $\rho$ behaves as
 \be\label{rhorstar}
 \rho \sim -r_*
 \ee
 in this region. This gives us the solution (\ref{R0sol}) which  behaves
as
 \be
 R^{(0)} \sim e^{-i\omega_0 r_*},
 \ee
for $r\rightarrow r_+$. This is  the ingoing mode propagating into
the horizon.  For the $n$-th radial eigenfunction, one can easily
show by induction
 \be
 R^{(n)} \sim e^{-i \omega_n r_*}, ~~{\rm as}~ r_* \to -\infty.
 \ee

Finally, we observe that $R^{(0)}(0)=0$ at infinity $\rho\rightarrow
0 ~( r_* \to \infty)$, which shows that it is not the outgoing wave
at infinity  but satisfies the Dirichlet boundary condition as like
at the infinity of AdS spacetime.  Moreover, the first radial
eigenfunction $R^{(1)}(\rho)$ can be explicitly constructed as
 \be
 R^{(1)}(\rho) =
  -2iC\omega_0\cosh(\kappa\rho)\Big[\sinh\left(\kappa\rho\right)\Big]^{i\frac{\omega_0}{\kappa}},
 \ee
which also satisfies the Dirichlet boundary condition at infinity.
One can easily show that the $n$-th radial eigenfunction $
R^{(n)}(\rho)$ behaves as the same way as $R^{(1)}(\rho)$ likewise.

\section{QNFs around near-extremal RN black hole}

We know that the literature of quasinormal modes for the RN black
hole is vast.   It is unclear that   the purely imaginary QNFs
(\ref{impq}) capture what kind of RN black hole, even though they
have obtained from the near-region and low-energy approximations of
a scalar wave equation around the RN black hole.  This requires to
know an appropriate  picture of a scalar perturbation around a
specified RN black hole. We wish to show that  {\it the specified RN
black hole is exactly the near-horizon region of a  near-extremal RN
black hole}.

For this purpose, we briefly mention  a known way to obtain the QNFs
of the near-extremal RN  black hole~\cite{Chen:2012zn}.  Then  we
compare the previous results with the ones obtained from the
near-extremal RN black hole.  By taking the near-horizon and the
near-extremal limits,
 \be
 r\rightarrow Q+\tilde{\rho},~~~M\rightarrow Q+r_0,
 \ee
the RN metric (\ref{MF}) can be rewritten  by two parameters
$r_0=\sqrt{2Q(M-Q)}$ and $Q$ as
 \be\label{nebh}
 ds^2_{\rm
 NHNE}=-\frac{\tilde{\rho}^2-r_0^2}{Q^2}dt^2+\frac{Q^2}{\tilde{\rho}^2-r_0^2}d\tilde{\rho}^2+Q^2d\Omega^2_2,
 \ee
 whose geometry is given by AdS$_2\times S^2$.  It is pointed out that (\ref{nebh}) with $r_0=0$
describes the near-horizon region of the extremal RN black hole
which was known to be the Bertotti-Robinson geometry.
 We note that the geometry (\ref{nebh}) describes the near-horizon region of a near-extremal RN black hole only,
but not the whole geometry of a near-extremal black hole because the
geometry at infinity is an  asymptotically flat spacetime. This
point is important to understand why QNFs (\ref{impq})  are purely
imaginary. Here $\tilde{\rho}$ describes outside the black hole
($\rho \in
 [r_0,\infty]$), which is an extended geometry of the near-horizon region.
For this near-extremal RN black hole, the surface gravity is given
by
 \be\label{surg1}
\tilde{\kappa}=\frac{r_0}{Q^2}.
 \ee
Here we note that the surface gravity (\ref{surg1}) is obtained
directly from the near-extremal solution (\ref{nebh}), while  one
can approximate the surface gravity (\ref{surg}) to give
 \be
 \kappa \approx \tilde{\kappa }\Big[1-{{\cal O}\Big(\frac{2r_0}{M}\Big)}\Big].
 \ee
We observe that for the near-extremal RN black hole, its surface
gravity is given by $\tilde{\kappa}$.  Then, the Klein-Gordon
equation (\ref{rhoKG}) with the ansatz (\ref{ansatz}) can be written
by
 \be
 \frac{d}{d\tilde{\rho}}\left((\tilde{\rho}^2-r_0^2)\frac{d}{d\tilde{\rho}}\right)R(\tilde{\rho})
 +\left(\frac{\omega^2Q^4}{\tilde{\rho}^2-r_0^2}-l(l+1)\right)R(\tilde{\rho})=0.
 \ee
Introducing the tortoise coordinate defined by
 \be
 \rho_*=\frac{1}{2\bar\kappa}\ln\left(\frac{\tilde{\rho}+r_0}{\tilde{\rho}-r_0}\right),~~\rho_*\in
 [\infty,0],
 \ee
the Klein-Gordon equation becomes the Schr\"odinger-type equation
 \be \label{sch4}
 \frac{d^2}{d\rho^2_*}R(\rho_*) + \Big[\omega^2-V_{\rm NE}(\tilde{\rho})\Big]R(\rho_*)=0,
 \ee
where the near-extremal RN potential is given by
 \be\label{exrn}
 V_{\rm NE}(\tilde{\rho})=\frac{l(l+1)(\tilde{\rho}^2-r_0^2)}{Q^4}.
 \ee
Moreover, solving the tortoise coordinate in terms of $\tilde{\rho}$
as
 \be
 \tilde{\rho}=r_0\coth(\tilde{\kappa} \rho_*),
 \ee
one can easily show that the near-horizon and near-extremal RN
potential leads to
 \be \label{neblp}
 V_{\rm NHNE}(\rho_*)=\frac{l(l+1)\tilde{\kappa}^2}{\sinh^2(\tilde{\kappa} \rho_*)},
 \ee
which is the exactly same form of $V_{\rm HCS}(\rho)$ in Eq.
(\ref{pot3}) when replacing $\tilde{\kappa}$ and $\rho_*$ by
$\kappa$ and $\rho$. See Fig. 1-(b) for $V_{\rm NE}(\rho_*)$.   Here
$\rho_*$ mimics exactly $\rho$ for describing the region outside the
event horizon. This means that our previous results is valid for the
near-extremal RN black hole only where the surface gravity of the RN
black hole $\kappa$ is replaced by the surface gravity of the
near-extremal RN black hole $(\kappa \to\tilde{\kappa})$.

As obtained in~\cite{Chen:2012zn}, the QNFs of a massive charged
scalar with mass $m$ and charge $q$ around the near-horizon region
of a  near-extremal RN black hole is given by
 \be \label{tqnfs}
 \tilde{\omega}_n=-\tilde{\kappa} b-i\tilde\kappa\left(n+\frac{1}{2}\right),
 \ee
where  the parameter $b$ is given by
 \be
 b=\sqrt{(q^2-m^2)Q^2-\left(l+\frac{1}{2}\right)^2}.
 \ee
In this work, since we are considering  the uncharged massless
scalar field, the parameter $b$ takes the purely imaginary value
 \be
 b=i\left(l+\frac{1}{2}\right).
 \ee
Plugging $b$ into (\ref{tqnfs}), one  obtains the QNFs
\be\label{neqnfs}
 \tilde{\omega}_n=-i\tilde\kappa\Big[n+l+1\Big],
 \ee
 for the scalar propagating on  the near-extremal RN black hole.
 Also, there are
other works which support our results. Kim and Oh~\cite{Kim:2001ev}
have computed QNFs ($\omega=-i\tilde{\kappa}(n+1)$) from the scalar
equation around the AdS$_2$ black hole obtained by making the
dimensional reduction of the Einstein-Maxwell theory on $S^2$.
Because of the dimensional reduction on $S^2$, the angular quantum
number $l$ was missed in $\omega$. Hod~\cite{Hod:2010hw} has shown
that QNFs of a  massive charged scalar propagating around the
near-extremal RN black hole are  given by (\ref{neqnfs}) in the
massless and uncharged cases.

In Appendix, we show explicitly that purely imaginary QNFs
(\ref{neqnfs}) could be obtained as (\ref{dqnfs})  by solving the
scalar wave equation (\ref{sch4}) around the near-horiozn region of
a near-extremal RN black hole directly. Furthermore, it is proven
that quasinormal modes as the solution to the scalar wave equation
around the near-horizon region of a near-extremal RN black hole
satisfy the ingoing-boundary condition at the horizon and Dirichlet
boundary condition at infinity.

The above states clearly that the QNFs of the scalar perturbation
around the RN black hole with  the hidden conformal symmetry are
those obtained from the scalar perturbation around the near-horizon
region of a near-extremal RN black hole. In this context, our QNFs
(\ref{impq}) are the same as (\ref{neqnfs}) obtained from the scalar
perturbation around the near-horiozn region of a near-extremal RN
black hole (\ref{nebh}) whose geometry is AdS$_2 \times S^2$. This
productive geometry describes effectively the two-dimensional black
hole whose quasinormal frequencies are usually given by purely
imaginary QNFs~\cite{Myung:2012cb}.

\section{Potential Picture}
\begin{figure}[t!]
   \centering
   \includegraphics{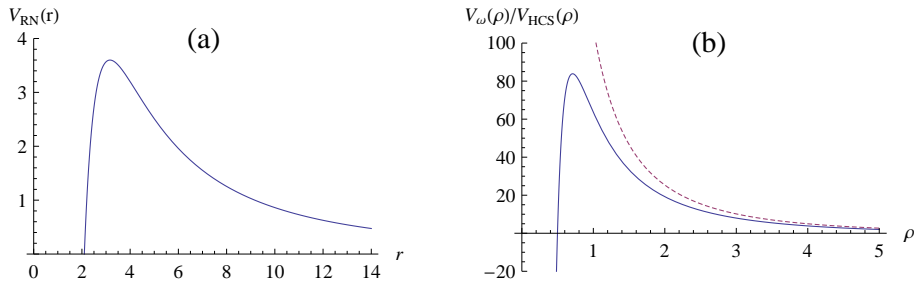}
\caption{Potential pictures for $l=10$: (a) the RN potential $V_{\rm
RN}(r)$ with $M=1.10$, $Q=0.46$, (b) the $\omega$-dependent
potential $V_\omega(\rho)$ in solid curve with $\omega=1$, $r_0=1$,
$r_-=0.10$, and the HCS potential $V_{\rm HCS}(\rho)$ in dashed
curve with $r_0=1$. We note that the potential $V_{\rm
NHNE}(\rho_*)$ is the same graph as in $V_{\rm HCS}(\rho)$ when
replacing $\rho$ by $\rho_*$. } \label{fig.1}
\end{figure}

Now we wish to support the validity of the QNFs $\omega_n$ in Eq.
(\ref{impq}) by comparing the relevant potentials.  In Fig.1, we
depict  these potentials with $l=10,
r_0=1~(r_+=2.1,r_-=0.1,M=1.1,Q=0.46,\kappa=0.23)$ for $V_{\rm
RN}(r),~V_{\rm \omega}(\rho)$ with $\omega=1$,  $V_{\rm HCS}(\rho)$,
and $V_{\rm NE}(\rho_*)$.

First, as is shown in Fig.1-(a), $V_{\rm RN}(r)$ is the potential
barrier existing outside the event horizon ($r=r_+$) of the RN black
hole. It is hard to  find  the analytic expression for the QNFs
because of asymptotic behavior of $V_{\rm RN}\sim 1/r^2$ as $r\to
\infty$. Accordingly,  the literature of quasinormal modes QNFs  for
the RN black hole is vast.

Second, as depicted by a solid curve in Fig.1-(b), the
$\omega$-dependent potential $V_\omega(\rho)$ shows a negative
potential around $\rho=0~(r\to \infty)$. At this stage, it seems
that we do not know  the appearance  of the negative potential which
may induce the instability of the RN black hole. In order to
understand the negativeness of the potential, we take its limit of
$\rho\to 0$ as
 \be
 V^{\rho\to 0}_{\rm \omega}(\rho)
 =\omega^2-\omega^2r_+^4\Big(\frac{1}{\rho}+\frac{r_++r_-}{2r_+^2}\Big)^4
 +\frac{l(l+1)}{\rho^2},
 \ee
where the second term is responsible for the negativeness in the
solid curve in Fig.1-(b), while the last term makes the potential
positively infinite as drawn by the dashed curve in Fig.1-(b).
Hence, imposing the near-region and low-energy limits, one may
neglect the second term in favor of the last term, effectively
leading to the HCS-potential depicted as the dashed curve in
Fig.1-(b).

Third, concerning  the $V_{\rm HCS}(\rho)$ potential, we have to say
that this potential is not a genuine potential which is valid for
the whole  RN black hole but a form of the potential obtained by the
approximation (\ref{app3}) to develop the hidden conformal symmetry
near the event horizon.   As was shown in section 4, $V_{\rm
HCS}(\rho)$ coincides with  the potential $V_{\rm NHNE}(r_*)$ in
(\ref{neblp}) of the scalar perturbation around the near-horizon
region of a near-extremal RN black hole.

Finally, we mention that the graphs in Fig.1 are designed for a
non-extremal RN black hole of
$r_+=2.1,r_-=0.1,M=1.1,Q=0.46,\kappa=0.23$. Even for the
near-extremal RN black hole, the behaviors of $V_\omega(\rho)$ and
$V_{\rm HCS}(\rho)$  are similar to those in Fig.1-(b) because
$\rho$-coordinate was used to draw the pictures whose range is from
$\rho=\infty$ (event horizon) to $\rho=0$(infinity).

\section{Conclusion}
In conclusion, we have derived the purely imaginary QNFs
(\ref{impq}) of the  RN black hole by making use of the hidden
conformal symmetry developed in the near-region  and low-energy
approximations of the massless Klein-Gordon equation.    We see that
the operator approach used to derive QNFs  has a limitation because
the $\omega$-dependent potential (\ref{pot2}) is just replaced  by
the HCS potential (\ref{pot3}), as is shown in Fig.1-(b). This means
that developing the hidden conformal symmetry in the near-region and
low-energy approximations  implies loosing the large $r$ behavior of
the potential in the whole RN black hole.  In deriving quasinormal
modes, thus, the outer-boundary condition at infinity was changed
from outgoing modes to Dirichlet boundary condition.

Consequently, the imaginary QNFs (\ref{impq}) based on the hidden
conformal symmetry are nothing but the quasinormal frequencies
(\ref{neqnfs}) of the scalar perturbation around the near-horizon
region of a near-extremal RN black hole whose geometry is described
by AdS$_2 \times S^2$. Since this geometry is not the whole geometry
of near-extremal RN black hole whose geometry at infinity is
asymptotically flat, QNFs are purely imaginary, in comparison  with
the complex value~\cite{Berti:2009kk}.

\section*{Appendix: A computation of QNFs around the near-extremal RN black hole}

In  appendix, we show that a scalar propagating around the
near-horizon region of a near-extremal RN black hole has purely
imaginary QNFs. Let us start with the Schr\"odinger equation
(\ref{sch4}) with the near-horizon and near-extremal RN potential
(\ref{neblp}). Introducing a new variable~\cite{Cardoso:2001hn}
 \be
 x=\frac{1}{\cosh^2(\tilde{\kappa}\rho_*)},~~x\in [0,1],
 \ee
Eq.~(\ref{sch4}) can be written as
 \be\label{nebhde}
 x(1-x)\frac{d^2}{dx^2}R+\left(1-\frac{3}{2}x\right)\frac{d}{dx}R
 +\left(\frac{\omega^2}{4\tilde{\kappa}^2x}-\frac{l(l+1)}{4(1-x)}\right)R=0.
 \ee
By changing to a new wave function $y$ through
 \be
 R=x^{-\frac{i\omega}{2\tilde{\kappa}}}(1-x)^{-\frac{l}{2}}y,
 \ee
Eq.~(\ref{nebhde}) can be transformed  into a standard
hypergeometric equation~\cite{Cordero:2012je}
 \be\label{hyperg}
 x(1-x)y''+[c-(a+b+1)x]y'-aby=0
 \ee
with
 \ba \label{abc}
 a=-\frac{i\omega}{2\tilde\kappa}-\frac{l}{2}+\frac{1}{2},~~
 b=-\frac{i\omega}{2\tilde\kappa}-\frac{l}{2},~~
 c=1-\frac{i\omega}{\tilde\kappa}.
 \ea
In order to obtain the quasinormal modes,  we have to check whether
or not solutions of Eq.~(\ref{nebhde}) satisfy the boundary
conditions: ingoing waves near the horizon ($x=0$) and zero
(Dirichlet  condition) at infinity of $x=1$.  Taking into account
$e^{-i\omega t}$,  such a solution takes the form of
 \be \label{qnmhg}
 R=x^{-\frac{i\omega}{2\tilde{\kappa}}}(1-x)^{-\frac{l}{2}}{}_2F_1[a,b,c;x],
 \ee
 where ${}_2F_1[a,b,c;x]$ is a standard hypergeometric function.
Imposing the boundary condition at infinity ($x\rightarrow
1,~\rho_*\rightarrow 0,~\tilde\rho\rightarrow\infty)$, we recall  a
property of the hypergeometric function
 \be
 \lim_{x\rightarrow 1}{}_2F_1[a,b,c;x]=\frac{\Gamma(c)\Gamma(c-a-b)}{\Gamma(c-a)\Gamma(c-b)}.
 \ee
The Dirichlet boundary condition can be achieved when choosing
 \be \label{con10}
 c-a=-n~{\rm or}~~c-b=-n
 \ee
with $n=0,~1,~2,\cdot\cdot\cdot$. Therefore, using (\ref{con10})
together with (\ref{abc}),  the QNFs are given by
 \be
 \omega_n=-i\tilde\kappa(2n+l+1),~~~\omega_n=-i\tilde\kappa(2n+l+2),
 \ee
which are combined to give a single expression of purely imaginary
QNFs
 \be
 \tilde{\omega}_n=-i\tilde\kappa(n+l+1). \label{dqnfs}
 \ee
This is the exactly same form as found in the QNFs (\ref{neqnfs}).

\section*{Acknowledgement}
We would like to thank B. Chen and C. -M. Chen for helpful
discussions. This work was supported by the National Research
Foundation of Korea (NRF) grant funded by the Korea government
(MEST) through the Center for Quantum Spacetime (CQUeST) of Sogang
University with grant number 2005-0049409. Y. S. Myung was also
supported by the National Research Foundation of Korea (NRF) grant
funded by the Korea government (MEST) (No.2012-R1A1A2A10040499).
Y.-J. Park was also supported by World Class University program
funded by the Ministry of Education, Science and Technology through
the National Research Foundation of Korea(No. R31-20002).

\end{document}